\documentstyle[aps,newdoc,epsf]{revtex}
\begin{document}
\epsfverbosetrue
\setcounter{tracingmulticols}{1}

\title{A model of linear chain submonolayer structures.\\
Application to Li/W(112) and Li/Mo(112)}

\author{F.~Bagehorn}
\address{Institut f\"{u}r Theoretische Physik, TU Dresden,
01062 Dresden, Germany}
\author{J.~Lorenc, Cz.~Oleksy 
\thanks{To be published in Surface Science, Corresponding author.
Fax: 48 71 214454; E-mail:  oleksy@ift.uni.wroc.pl}}
\address{Institute of Theoretical Physics, University of Wroc{\l}aw,
Plac Maksa Borna 9, 50-204 Wroc{\l}aw, Poland}

\maketitle
\begin{abstract}

We propose a lattice gas model to account for linear chain structures
adsorbed on (112) faces of tungsten and molybdenum. This model
includes a di\-po\-le-di\-po\-le interaction as well as a long-ran\-ge
indirect (oscillatory) interaction of the form
$\sim cos(2k_Fr+\varphi)/r$, where $k_F$ is the wavevector of electrons
at the corresponding Fermi surface and $\varphi$ is a phase shift. It
is assumed that the structures are stabilized by an attractive indirect
interaction along the chains.

We have explicitly demonstrated that the periodic ground states
strongly depend on a competition between the di\-po\-le-di\-po\-le 
and long-ran\-ge indirect interactions.

The effect of temperature in our model of linear chain structures is
studied within the mo\-le\-cu\-lar-field approximation. The numerical
results clearly show that for the di\-po\-le-di\-po\-le interaction
only , all long-per\-io\-dic linear chain phases are suppressed to low
temperatures while phases with periods  2, 3, and 4  dominate the phase
diagram. However, when the long-ran\-ge indirect interaction becomes
important, the long-per\-io\-dic linear chain phases start to fill up
the phase diagram and develop a high thermal stability.

We have chosen model parameters in order to reconstruct a sequence of
long-per\-io\-dic phases (for coverages less than 0.5) as observed
experimentally at $T=77K$ for Li/Mo(112) and Li/W(112). 
It would be interesting to verify our model and assumptions by
checking experimentally the corresponding phase diagrams.
\end{abstract}

\widetext
\begin{multicols}{2}
\section{Introduction}
Chemisorption on metal surfaces has been attracting a great deal of
attention for over three decades (for a review see, for example,
\cite{ref1}-\cite{ref6} and references cited therein). Its nature,
however, despite quite involved experimental techniques and a number of
data as well as many theoretical attempts is far from being well
understood.  One of the recent challenging problem has been concerned
with structures and phase transitions in metal submonolayers adsorbed
on (112) faces of tungsten and molybdenum \cite{ref1,ref4}. In
particular, it was found from LEED experiments that many alkaline,
al\-ka\-li\-ne-e\-arth, and ra\-re-e\-arth elements adsorbed on these
substrates form ordered structures which, for low coverages, consist of
linear chains of adatoms being perpendicular to the atomic troughs at
the surface. The chains of adatoms are often far apart from one another
in the direction along the atomic troughs (up to 9 lattice constants of
the substrate) thus forming {\it long-per\-io\-dic} (linear) chain
structures. Moreover, it turns out that the structures possess a high
thermal stability \cite{ref4}.

At higher coverages, however, submonolayer structures become much more
complex for, in addition to coherent structures, there are 
one-di\-men\-sio\-nal incoherent or stressed submonolayer structures 
\cite{ref4}.

It is clear, that lateral interactions of adatoms are most important in
determining the properties of adsorbed overlayers like their structures,
phase transitions, thermal stability, etc. \cite{ref1}. These
interactions at (112) surfaces of bcc metals (W, Mo) are believed to be
due to a repulsive (isotropic) di\-po\-le-di\-po\-le interaction between
adatoms with  dipole moments, as well as to a long-ran\-ge indirect
(anisotropic) interaction via substrate electrons \cite{ref1,ref4}.
The delicate balance between the two interactions seems to account, at
least for low coverages, for the observed properties of adsorbed linear
chain submonolayers. So far, only qualitative (geometrical) arguments
have been put forward in an attempt to understand to what extent the
indirect electronic interaction might be responsible for the observed
chain structures \cite{ref4}, \cite{ref7}-\cite{ref8}.

The purpose of this paper is to study how a {\it competition} between
the (repulsive) di\-po\-le-di\-po\-le and (oscillatory) long-ran\-ge
indirect electronic interactions could influence ordered submonolayer
structures as well as their equilibrium thermodynamic properties (phase
diagram). This is done within a plausible two-di\-men\-sio\-nal lattice
gas model which might describe linear chain structures for coverages up
to 0.5 .

The model is described in detail in section~\ref{s2}. In
section~\ref{s3} we calculate numerically the ground states ($T=0K$) and
discuss their dependence on the model parameters. The mean-field
approximation is used  in section~\ref{s4} to calculate the
corresponding phase diagrams. A possible application
to describe low coverage linear chain submonolayers of Li absorbed on
W(112) and Mo(112) has been presented in subsection~\ref{s4s2}.  
Finally, section~\ref{s5} contains a general discussion and conclusions.

\section{The model}\label{s2}
\begin{figure}[h]
\epsfxsize=9cm
\epsfbox{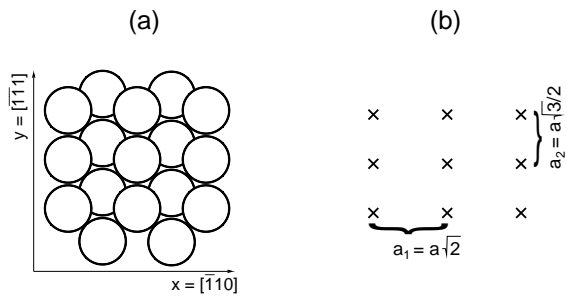}
\caption{ (a) A schematic view of the (112) surface of bcc\\
metals. (b) The corresponding (rectangular) lattice of ad-\\
sorption sites shown by the crosses  ($a=3.16\AA$ for W and\\
$a=3.14\AA$ for Mo)}.
\label{scheme}
\end{figure}

It is generally believed, that chemisorbed atoms occupy, at low coverage
at least, only preferred sites of a lattice commensurate with the
underlying substrate lattice. Here, we consider the (112) surface of
bcc metals (W or Mo) and the corresponding lattice of adsorption sites
\cite{ref4} (see fig.~\ref{scheme}).

It turns out, that lattice gas models are quite useful in studying
overlayer structures and their properties (e.g. refs \cite{ref2,ref9}).
As usual, a lattice
gas model introduces the occupation variable $\;n_{\vec{r}}\;$ at each
adsorption site $\vec{r}$: $n_{\vec{r}}=1$ if the $\vec{r}$th site is
occupied by an adatom and $n_{\vec{r}}=0$, if not. Then, within the
grand canonical ensemble, the Hamiltonian is defined as
\begin{equation} \label{hamiltonian}
{\cal H}=\frac{1}{2}\sum_{\vec{r}\in L}\sum_{\vec{r}^\prime\in L \atop
\vec{r}^\prime\neq\vec{r}} V(\vec{r}-\vec{r}^\prime)n_{\vec{r}}
n_{\vec{r}^\prime}-\mu\sum_{\vec{r}\in L}n_{\vec{r}} \;\;\;\;\mbox{,}
\end{equation}   
where L denotes the rectangular lattice of N$\times$N adsorption sites
with periodic boundary conditions and $\mu$ is the chemical potential
(shifted, here, by the binding energy of an adatom to the substrate).
Of course, $\mu$ controls the adatom coverage
\begin{equation}
\theta =\frac{1}{N^2}\sum_{\vec{r}\in L}\langle\: n_{\vec{r}}\:\rangle
\end{equation}
with $\langle\:\ldots\:\rangle$ meaning a thermodynamic average. We are
interested in a lattice gas model describing linear chain structures as
observed experimentally for low coverages ($\theta <0.5$) and, 
herefore, we can neglect ma\-ny-bo\-dy interactions. Since, even in this
case, the reliable first-prin\-cip\-le calculation of effective lateral
interactions is very difficult, following a detailed discussion in
ref.~\cite{ref4}, we have considered a pairwise adsorbate interaction
$V(\vec{r})$ consisting of electrostatic and indirect interactions.
Thus, we have assumed
\begin{equation} \label{potential}
V(\vec{r})=\frac{2d^2}{|\vec{r}|^3}+\left\{
\begin{array}{lr}
A\cos (2k_F |y|+\varphi )\frac{1}{|y|}\delta (x,0)&\;{\rm for}\;
y\neq 0\\
\\
E_b \delta (|x|,a_1)&\;{\rm for}\; y=0
\end{array} \right. \;\;\;\;\mbox{,}
\end{equation}  
where $\vec{r}=(x,y)$ and $\delta$ stands for  Kronecker's symbol
(for notation see also fig.~1). These interactions are of
essential importance in forming linear chain structures of alkalis,
al\-ka\-li\-ne-earths, and ra\-re-earths adsorbed on the (112) surface
of W or Mo and they have the following meaning:
\renewcommand{\labelenumi}{(\roman{enumi})}
\begin{enumerate}
\item The first component in eq.~(\ref{potential}),
$\; 2d^2/|\vec{r}|^3\;$, describes a repulsive di\-po\-le-di\-po\-le
interaction between two identical adatoms adsorbed at $\vec{r}$ and
$\vec{r^\prime}$, respectively (see, for example, refs \cite{ref4} and
\cite{ref6}). The dipole moment of an adsorption bond, $d$, is usually
estimated from the Helmholtz formula. Here, we neglect all
depolarization effects due to ma\-ny-bo\-dy interactions and this seems
to be justified for $\theta<0.5$.
\item The following component for $y\neq 0$,
$\; A\cos (2k_F|y|+\varphi )\frac{1}{|y|}\delta (x,0)\;$,
represents an asymptotic part of the indirect interaction between
adatoms via conduction electrons of the substrate (all relevant
literature on this interaction could be found in refs \cite{ref4}
and \cite{ref10}). This particular interaction is highly anisotropic
and is closely related to the existence of nearly flattened segments
of the Fermi surface being perpendicular to the $[\:\overline{1}
\overline{1}1\: ]$ axis directed along the atomic troughs of the
substrate. An amplitude $A$ and a phase $\varphi$ can be treated as
phenomenological parameters and $k_F$ denotes a wavevector of electrons
at the Fermi surface. In our model we assume, following 
refs \cite{ref11,ref12}, that $k_F = 0.41\AA^{-1}$ 
($k_F = 0.47\AA^{-1}$) for the tungsten (molybdenum) substrate. 
It is needless to say, that this
indirect (oscillatory) interaction is long-ran\-ged ($\sim |y|^{-1}$)
and it could influence, as usually expected, properties of the absorbed
submonolayers.
We have neglected all contributions to the indirect interaction
related to hole segments of the Fermi surface \cite{ref4,ref13}.
\item Finally, for $y=0$ the $\;E_b \delta (|x|,a_1)\;$ term
has been introduced to eq.(\ref{potential})
to facilitate the formation of
linear chains of adatoms along the $x$-direction. This is a part of
an {\it attractive} indirect interaction between adatoms along a chain
and we assume that
$E_b=- 0.05 {\rm eV}$ at the
nearest-neighbour distance $a_1$ \cite{ref4}.
\end{enumerate}
\indent The model interactions, eq. (3), requires, however, some
justification.
\par
\noindent First, the indirect, substrate mediated interaction (ii)
reflects only the leading "effective" dimensionality of the electronic
states (e.g. refs \cite{ref10},\cite{ref14}-\cite{ref15})
mediating the interaction in the direction perpendicular to the chains.
Here, we neglect a usual part of the indirect interaction decaying
like $\frac{1}{r^5}$ and associated with the "spherical" segments of the
Fermi surface \cite{ref4,ref16}. Of course, $r$ denotes a distance
between two adatoms.
\par
\noindent Secondly, we would like to comment on what role surface
electronic states may play in the indirect interaction. It is known,
that they could contribute to this interaction provided the Fermi energy
level intersects a band of the surface electronic states. Then,
depending on the effective dimensionality of the surface states, the
indirect interaction becomes proportional either to $\frac{1}{r}$
(quasi--one--dimensional surface states \cite{ref10},
\cite{ref14}-\cite{ref15}) or to $\frac{1}{r^2}$ (two--dimensional
surface states \cite{ref4}, \cite{ref15}-\cite{ref16}). A period of the
oscillations will be determined by a wavevector at the Fermi surface 
corresponding to the surface electronic states. Recently, the field and 
photofield emission experiments on the clean W(112) surface have
revealed the existence of surface states approximately 0.3$eV$ below the
Fermi energy \cite{ref17}-\cite{ref18}. To our knowledge, however, there
is so far {\it no} experimental evidence that surface states mediate the
indirect interaction in the case of the linear chain submonolayers on
the (112) surface of W(Mo). Consequently, we have neglected a would be
contribution to eq. (3) from the surface states.
\par
\noindent Thirdly, the model interaction (iii) could be thought of as a
usual substrate mediated interaction in the form
$\sim \cos(2k_{F}x + \phi)\delta(y,0)/|x|^5$ (see, for example, 
refs ~\cite{ref16}, \cite{ref19}-\cite{ref22}). In order to describe
the linear chain structures we might assume that $\phi$ causes this
indirect interaction to be attractive at the nearest--neighbour distance
$a_{1}$ and to decay rapidly with distance. We cannot also exclude a
possibility that (iii) might be due to a virtual elastic distortion of
the substrate surface \cite{ref23}.

At this stage it is important to notice, that although the interaction
described by eq. (\ref{potential}) is long ranged, we may restrict
ourselves to an extended but finite range of interaction, $R$. A more
detailed discussion will be presented in the following section.

\section{Properties of the ground states}
\label{s3}
A search for all the lowest energy configurations at $T=0{\rm K}$
for the model described by eqs (\ref{hamiltonian}) and
(\ref{potential}) presents a formidable problem. Even in a simpler
case of the di\-po\-le-di\-po\-le interaction only, as far as we know, 
it has not been possible up to now to find all the ground states. (For
a review of the existing theoretical and numerical results concerning
the ground states in two-di\-men\-sio\-nal lattices see, for example,
refs \cite{ref24}-\cite{ref26} and the literature cited therein.).
However, let us note that the ordered phases found experimentally at
$T=77 {\rm K}$, as we have already discussed, are often composed of
parallel linear chains, at least for low coverages ($\theta <0.5$).
This suggests, that a search for the minimal grand canonical
ensemble energy configurations (the ground states) could be performed
in two steps. First, we find the effective Hamiltonian describing
interactions between linear chains and only then we look for the
ground states.

\subsection{The effective Hamiltonian}
Making use of the periodicity along the chains, we can write down the
Hamiltonian as an effective Hamiltonian of a one-di\-men\-sio\-nal
lattice gas type model, which describes the interaction between linear
chains (simply referred to as chains, hereafter). This Hamiltonian has
the form
\begin{equation} \label{row4}
{\cal H}_{\rm eff} = \frac{1}{2}\sum_j\sum_{l(\neq 0)}V(l)\sigma_j
\sigma_{j+l} -(\mu +\varepsilon_c )\sum_j\sigma_j \;\;\;\;\mbox{,}
\end{equation}
where $\sigma_j$ denotes chain variable with $\sigma_j =1$ or 0
depending on whether there is a chain on a site j along the y direction 
or not. The chain--chain interaction energy at a distance $a_2|l|$ reads
\addtocounter{equation}{-1}
\begin{mathletters}
\begin{equation} \label{row4a}
V(l)=\sum_{k=-\infty}^{+\infty}V_d(k,l)+V_{ind}(l) \;\;\; ,
\end{equation} 
where
\[ V_d(k,l)=\frac{2d^2}{[k^2a_1^2+l^2a_2^2]^{3/2}} \;\;\; ,\]
\[ V_{ind}(l)=A\frac{\cos (2k_Fa_2|l|+\varphi )}{a_2|l|}\;\;\; .\]
The binding energy per chain can be written as
\begin{equation}\label{row4b}
\varepsilon_c =-\left(\frac{2d^2}{a_1^3}\:\zeta (3)+E_b\right)\;\;\; .
\end{equation}
\end{mathletters}
Here, $\zeta (3)=1.202056$\dots is the Riemann ze\-ta-func\-tion and
j, l, and k are integers.

It is well known, that the lattice gas effective Hamiltonian,
${\cal H}_{\rm eff}$, is analogous to the Hamiltonian of the Ising
model in an applied field. For a class of long-ran\-ge
interactions (positive, convex, etc.) it has been possible to find
in a one-di\-men\-sio\-nal case all the ground states for any rational
$0<\frac{q}{p}<1$ ($p$, $q$ are integers with no common multipliers)
\cite{ref26}. Moreover, the $(\theta , \mu)$ phase diagram represents
the complete devil's staircase with a rather involved fractal behaviour
\cite{ref26}-\cite{ref29}. Unfortunately, the effective Hamiltonian,
${\cal H}_{\rm eff}$, contains the indirect (oscillatory) interaction
and, therefore, we are not able to find the ground states rigorously.
Also, the results of  refs \cite{ref30,ref31} do not seem to be
applicable because the considered interaction is truly long-ran\-ged.

\subsection{A search for the ground states}
As usual, the long-ran\-ge nature of the effective interaction between
chains described by eq. (\ref{row4a}) poses a problem. It is
easy to see that the interaction is dominated at large distances
by the (oscillatory)
indirect term and one has to sum up the infinite series 
(see eq. (\ref{row4})) for any would-be minimal energy configuration of
$\sigma$'s. Instead, we approximate the problem (see also refs
\cite{ref32} and \cite{ref25}) by assuming that there is a sufficiently
large range of the adatom - adatom interaction, say, $R\: a_2$,
beyond which a configuration under study is replaced by the average
one having the same coverage $\theta$. This average chain structure
consists of equidistant "chains" with a spacing $a_0=a_2/\theta$.
Moreover, adatoms are uniformly distributed along these "chains" with
a density $1/a_1$.

Finally, we restrict ourselves to periodic configurations of
$\sigma$'s, i.e. there is a period $p$ (positive integer) such that
$\sigma_{j+p}=\sigma_j$ for any integer $j$. These are called
$p$-periodic configurations of $\sigma$'s.

Now, the energy per chain of a given $p$-periodic configuration
$\sigma_0,\dots,\sigma_{p-1}$ can be written in the form
\end{multicols}
\noindent\makebox[87mm]{\hrulefill}
\begin{equation} \label{row5}
E[\sigma_0,\dots,\sigma_{p-1}]=\frac{1}{p}\sum_{j=0}^{p-1}
\sum_{l=1}^R\tilde{V}(l)\sigma_j\sigma_{mod_{p}(j+l)}-
\frac{1}{p}(\mu +\varepsilon_c)\sum_{j=0}^{p-1}\sigma_j +E_r(\theta)
\;\;\;\;\mbox{,}
\end{equation}
\hspace*{93mm}\hrulefill
\begin{multicols}{2}
where
\begin{equation} \label{row5a}
\tilde{V}(l)=\sum_{k= -k(l)}^{k(l)}V_d(k,l)+V_{\rm ind}(l)
\end{equation}
with $k(l)={\rm Int}(\frac{a_2}{a_1}\sqrt{R^2-l^2})$.

Here, the remainder of the energy per chain reads
\begin{eqnarray}\label{row5b}
E_r(\theta )&=&\theta^3\frac{4d^2}{a_1a_2^2}\left[\frac{\pi^2}{6}-
\sum_{n=1}^{{\rm Int}(R\theta )}\frac{\sqrt{1-\left(\frac{n}{R
\theta}\right)^2}}{n^2}\right] + \nonumber\\
& & \nonumber\\
& &\mbox{}+\theta^2\frac{A}{a_2}\left[
f_1(\tilde{\alpha})\cos\varphi +f_2(\tilde{\alpha})\sin\varphi
+r(\alpha )\right]\;\;\;\;\mbox{,}
\end{eqnarray}
where
\[\theta =\frac{1}{p}\sum_{j=0}^{p-1}\sigma_j\;\;\;\;\mbox{,}\]
\[\alpha =\frac{2k_Fa_2}{\theta}\;\;\;\;\mbox{,}\]
\[f_1(\tilde{\alpha})=-\ln (2\sin\frac{\tilde{\alpha}}{2})\;\;\;\;
{\rm for}\;\tilde{\alpha}={\rm mod}_{2\pi}(\alpha )>0\;\;\;\;\mbox{,}\]
\[f_2(\tilde{\alpha})=\frac{\tilde{\alpha}-\pi}{2}\;\;\;\;\;\;\;\;\;
{\rm for}\;\tilde{\alpha}={\rm mod}_{2\pi}(\alpha )>0\;\;\;\;\mbox{,}\]
\[ r(\alpha )=\sum_{n=1}^{{\rm Int}(R\theta )}
\frac{\cos (n\alpha +\varphi -\pi )}{n}\;\;\;\;\mbox{.}\]
The first two components in eq. (\ref{row5}) depend on
$\sigma_0,\dots,\sigma_{p-1}$ and involve exact summations. The
remainder of the energy per chain, however, depends only on $\theta$ as
it should have been expected. It is interesting to note that this model
removes the particle-hole symmetry \cite{ref2}.

In the following, we propose a numerical procedure to find the
(periodic) ground states. First, by using the bit representation of
integers from the interval $[2^{p-1},2^p-1]$ we generate numerically
all $p$-periodic configurations of $\sigma$'s with $p=1,\dots,
p_{max}$. The number of configurations could then be reduced by
making use of the translational and/or inversion symmetries.
Secondly, we check explicitly which configurations of $\sigma$'s
afford the minimal value to the corresponding $E[\sigma_0,\ldots,
\sigma_{p-1}]$ ($p=1,\ldots,p_{max}$). Of course, the results
depend on model parameters, such as $d, A, \varphi, k_F,  R$ and 
$p_{max}$.
This will be discussed in the following subsection.

\subsection{The results}
\label{s3s3}
To describe a competition between the di\-po\-le-di\-po\-le and indirect
interactions we require three independent parameters: 
$\;\frac{d^2}{A}, \varphi$, and  $k_F$.  This results from
eq.~(\ref{row5}) where $E[\sigma_{0},\ldots,\sigma_{p-1}]$ and
$\mu + \epsilon_c$ are rescaled by a factor of $A^{-1}$, thus changing
effectively only the stability intervals of the ground states but not 
the ground states themselves. Moreover, our approximation concerning
the energy per chain of a given p-periodic configuration of
$\sigma 's$ (eq. (\ref{row5})) and the numerical procedure introduce
two more parameters: $R$ and $p_{max}$. In principle, calculations
should be performed for very large $R$ (the larger the better) and for
$\;p_{max}\;$  going to infinity to allow for any periodic structure. In
practice, we have carried out calculations for $R=150$ and $p_{max}=23$.
This is a compromise between the computing time and more precise
results. Our numerical tests show that relative changes of energy at
$Ra_2=150a_2$ (in comparison with $Ra_2=1600a_2$) are less than one
percent for a vast majority of $p$-periodic configurations of
$\sigma's$   $(p=1,\ldots,p_{max}=23)$. Also, the results
for larger values of $\;p_{max}\;$ change the results only quantatively.
We believe, that to understand the role the indirect interaction plays
in determining the ground states, no additional significant insight can
be achieved by extending the numerical
computations to higher values of $\;p_{max}\;$.

We have applied our numerical procedure to determine the ground states
for a number of different parameters: $\;\frac{d^2}{A}, \varphi$, and
$k_F$. Some of the results are presented in tables~\ref{tab1}-\ref{tab2}
and fig.~2. The tables contain sequences of 
ground states. Here we use standard notation for the ground state, i.e.
$q/p$, which means $q$ chains in a unit cell of $p$ sites along the y  
direction (fig. 1). At the same time $\theta = \frac{q}{p}$ and in the
present analysis we are not interested in what are the actual
structures of the ground states. Also, it turned out to be convenient
to express the strength of the indirect interaction, $A$, in units
of $-0.137 eV\AA$ (see refs \cite{ref6} and \cite{ref16}). 

In fig. ~2 we have shown explicitly examples of stability intervals of
the corresponding ground states with respect to a reduced chemical
potential $\mu/\mu_0$ ($\mu_0=k_B T, T=100 K$). Note that only halves
of the stability intervals for $1/2$  and $2/4$ have
been depicted. As usual, the coverage as a function of reduced chemical
potential forms the so-called "staircase".

Now, we shall discuss the presented results paying a particular emphasis
to the role which the indirect interaction plays in determining the
ground states.

\vspace{10mm}
\par\noindent
i) {\em Dependence on} $\;\;\frac{d^2}{A}$

\vspace{4mm}
This parameter is a measure of a competition between the
di\-po\-le-di\-po\-le and indirect interactions. The results shown in
table~\ref{tab1} and fig.2 indicate that by decreasing $\frac{d^2}{A}$
(or by increasing the strength of the indirect
interaction) one causes the following changes to the ground states:
\begin{itemize}
\item For very large values of $\;\frac{d^2}{A}\;$ we practically obtain
the complete devil's staircase which is shown in fig.~2(a) (see also
refs \cite{ref26}-\cite{ref29}).
\item For large values of $\;\frac{d^2}{A}\;$ many narrow
long-per\-io\-dic ground states (most of which correspond to higher
coverages close to $0.5$) disappear from the staircase.
The remaining ground states usually change their stability intervals
and might transform into the ground states (with a close coverage) (see
fig.~2(b)).
\item For smaller values of $\;\frac{d^2}{A}\;$ the narrow
long-per\-io\-dic ground states (corresponding to low coverages)
continue to disappear and we are left with few ground states some of
which might be the long periodic ones (see fig.~2(c)). There can even be
one ground state (for example, $7/20$ for $d=0.3, A=1$, and other
parameters as in fig.~2). This qualitative argument is supported by our
calculations for other values of $\varphi$ and $k_F$. For example, it
also applies to $\varphi=0$ and $\varphi=1.6\pi$ with $k_F=0.41\AA^{-1}$
(see table~\ref{tab1}) but $\;\frac{d^2}{A}\;$ must be larger than $2.25$.
\end{itemize}
At this point we would like to note that in the case of a sufficiently
strong indirect interaction for $0.9\pi \le \varphi \le 1.5\pi$ and 
$0.40\AA^{-1} \le k_F \le 0.50\AA^{-1}$ there is only a coexistence of 
the low-den\-si\-ty ($\theta=0$) and high-den\-si\-ty ($\theta=1$)
disordered ground states. This means that an attraction effectively
prevails between chains, thus leading to formation of
two-di\-men\-sio\-nal islands. This particular result, however, seems
not to be accounted for by our model of linear chains (see
section~\ref{s3}).

\vspace{10mm}
\par\noindent
ii) {\em Dependence on} $\;\;\varphi \;$ and $\; k_F$

\vspace{4mm}
It is easy to see from tables~\ref{tab1}-\ref{tab2} that the ground
states depend on $\varphi$ and $k_F$ in a crucial way. The phase,
$\varphi$, simply shifts uniformly the locations of minima of the
corresponding indirect interaction thus allowing for the ground states
with possible new periodicities or$/$and changed stability intervals
(in $\mu$). The wavevector, $k_F$, distorts uniformly the minima
locations and a formation of  the ground states is in this case even
more complex. Indeed, the results presented in table~\ref{tab2} indicate
that for $\varphi=0$ each sequence of the ground states consists of the
ground states having one (two) long-ran\-ge period(s).  Within our model
this observation could be understood qualitatively by noting, that for a
given $k_F$, the actual ground state $q/p$ wins the energy competition
between two terms:  $E[\sigma_{0},\ldots,\sigma_{p-1}] -E_r(\theta)$
involving a structure of a would-be ground state (i.e.
$\sigma_{0},\ldots,\sigma_{p-1}$) and $E_r(\theta)$, where
$\theta= \frac{1}{p}\sum_{j=0}^{p-1}{\sigma_{j}}$ is such
that $\tilde{\alpha}$ is close to $0$ or $2\pi$ 
(see eqs (\ref{row5}), (\ref{row5a}), and (\ref{row5b})). This argument
seems to be valid also for $\varphi \in (0,\frac{\pi}{2}]$ and $\varphi
\in [\frac{3\pi}{2},2\pi)$ despite the fact that the ground states start
to gain on their energies. And this gain for $\varphi\in (\frac{\pi}{2},
\frac{3\pi}{2})$ is even more dramatic for now there are several new
ground states. Apart from long-per\-io\-dic ground states there are
also such ones as, for example, $1/2,\; 1/3\;$, and $\;1/4$
(see table~\ref{tab1}).

\section{Phase diagrams. Mean-field treatment}
\label{s4}

As a first step we adopt the mean-field approximation (a discussion
concerning the validity of this approach is postponed until the
following section) to describe (equilibrium) phase diagrams for linear
chains structures with the competing di\-po\-le-di\-po\-le and
long-ran\-ge indirect (oscillatory) interactions.  In the following, we
restrict ourselves to the effective Hamiltonian ${\cal H}_{eff}$,
eqs (\ref{row4}), (\ref{row4a}), and (\ref{row4b}) with a finite range
of the adatom - adatom interaction $R_{eff}$. The reason is that thermal
fluctuactions "screen out" the long-ran\-ge indirect interaction
\cite{ref4} and it is plausible to assume for a qualitative analysis, at
least, that $R_{eff}$ does not depend on temperature and the
di\-po\-le-di\-po\-le interaction can be neglected for distances larger
than  $R_{eff} a_2$.

Following the standard procedure (e.g. ref.~\cite{ref2}) we can write
the mean-field Hamiltonian in the form
\begin{equation} \label{row6}
{\cal H}_{eff}^{MF}=\sum_{i} {(V_i-\tilde\mu)\sigma_i}
-\frac{1}{2}\sum_{i}{V_i\langle\sigma_i\rangle}\;\;\;\;\mbox{,}
\end{equation}
where 
\addtocounter{equation}{-1}
\begin{mathletters}
\begin{equation}\label{row6a}
V_i=\sum_j{\tilde V(i-j)\langle\sigma_j\rangle}\;\;\;\;\mbox{,}
\end{equation}
and 
\begin{equation}\label{row6b}
\langle\sigma_i\rangle =\left[1+e^{\beta(V_i-\tilde\mu)}\right]^{-1}\; .
\end{equation}
\end{mathletters}

Again, $\sigma_j$ denotes the chain variable with $\sigma_j=0$ or $1$.
Moreover, $\tilde V(i-j)$ is given by eq. (\ref{row5a}) with $R$ replaced by
$R_{eff}$ and $\tilde\mu=\mu+\epsilon_c$ with $\epsilon_c$ defined via
eq. (\ref{row4b}).
 
First, we solve the self-con\-sis\-tent eqs (\ref{row6a}) and
(\ref{row6b}). This becomes possible to do, if we assume, that the
resulting structure is a $p$-periodic one,  $p=1, \ldots , p_{max}$.
Usually the solution will not be unique (i.e. several ordered structures
are possible). Secondly, having found the solution(s) we can calculate
the corresponding grand canonical ensemble potential(s) per chain
\begin{equation} 
\Omega(T,\tilde\mu)=-\frac{1}{p}\sum_{i=0}^{p-1}{
\left \{  ln\left[1+ e^{-\beta (V_i-\tilde\mu)} \right]
+\frac{1}{2} V_i\langle\sigma_i\rangle \right\} }\;\;\; .
\end{equation}

Hence, the stable structure corresponds to the minimal value of   
$\Omega(T,\tilde\mu)$, while the other structures are metastable or 
unstable ones.

In the computations we assume  $R_{eff}=36$ (i.e. $R_{eff} a_2\approx 
100\AA$) and $p_{max}=12$ according to the "lever rule" between
a computing time and more refined phase diagrams. It seems that
$p_{max}=12$ is sufficient to account for some linear chain structures
determined by the LEED technique \cite{ref4}.

\subsection{ Dipole-dipole interaction}

We start with a model of linear chain structures which order via the 
di\-po\-le-di\-po\-le interaction only (with an additional attraction
along the chains). This model might be relevant to adsorption on
furrowed surfaces, where the indirect interaction can be neglected
(small $A$ or lack of the corresponding flattened segments of the Fermi
surface of a substrate).

The results of the calculations are presented in fig.~3. This is a
temperature ($T$) versus reduced chemical potential ($\frac{\mu}{\mu_0},
\mu_0=k_BT, T=100 K$) phase diagram, where we have denoted equilibrium
(stable) phases by their ground states notation $q/p$. Note, that there
is one exception to the rule, i.e. the hig\-her-tem\-pe\-ra\-tu\-re
case, $4/10$, which has no the ground state equivalent. The phase
diagram is shown for $\mu > -0.72 \mu_0$ and this is due to our
restriction to $p_{max}=12$. The most interesting
result is the following: the thermal fluctuations suppress long periodic
phases and most of the phase diagram is dominated by the phases $1/2$,
$1/3$, and $1/4$.  This occurs despite the long-ran\-ge nature of the
di\-po\-le-di\-po\-le interaction.
(Note that a similar observation was reported in ref. \cite{ref33} for a
two-di\-men\-sio\-nal model with short-ran\-ge interactions).

\subsection{ Application to Li/W(112) and Li/Mo(112)}
\label{s4s2}

We claim that our model of linear chain structures might qualitatively
account for the adsorption of lithium atoms on the $(112)$ surfaces of W
and Mo \cite{ref4}, \cite{ref34}-\cite{ref35}. The adsorption bonds are
polarized and at a low coverage, $d=1.7D$ and $d=1.4D$ for the W and Mo
substrates, respectively. Also, we expect the indirect interaction to be
important (see section~\ref{s2}).  Finally, no "hard core" effects (a
relatively small atomic radius of a lithium atom) make the model more
realistic.

We have found two sets of the  model parameters which might be relevant
to Li/Mo(112) and Li/W(112) and the quantative features of the
corresponding mean-field phase diagrams are summarized in figs 4 and 5.
Both these figures have some common features. That is, the horizontal
hatching indicates coexistence regions between ordered (equilibrium)
phases denoted as in fig.~3. These are the so-called "mixed-pha\-ses"
regions. A dashed curve represents a second order phase transition curve
between the low density $(\theta =0)$ lattice gas phase and $1/2$ phase.
This curve meets, at some angle, two tangent first order phase
transition curves at a multicritical point P.

Now, a direct comparison of the phase transition schemes for Li/Mo(112)
\cite{ref34} and for Li/W(112)  \cite{ref35} with a low temperature vs
coverage sequence of transitions of fig.~4 and fig.~5, respectively,
shows a qualitative similarity.

In fig.~4, a low temperature scheme of phase transition  corresponds
quite well to Li/Mo(112) because the first order transitions occur
through forming islands of a phase in the "mixed-phase" region. It is
also interesting to note a high thermal stability of $1/4$ and $1/2$
phases at the corresponding stoichiometries.

The phase diagram presented in fig.~5, however, needs some more
explanation. The cross-hatch\-ing depictes a region where the results
do not seem to be reliable due to our restriction to $p_{max}=12$. We
have also found that at low temperatures and close to the disordered
phase ($\theta<0.05$) our numerical iteration procedure of solving
eqs~(\ref{row6a}) and (\ref{row6b}) turns out to be 
non-con\-ver\-gent (oscillatory behaviour). Therefore, we consider only
($\theta>0.05$). Even so, we recover (not too close to the disorder
phase) a low temperature sequence of $1/4$, $1/3$, and $1/2$ phases.
Of course, the  present theory cannot describe the phase transition
between $1/4$ and $1/3$ phases occuring via a kind of statistical
mixture of phases (cf. \cite{ref36,ref37}). Let us note that below
$\sim 80K$ there are long-per\-io\-dic structures (i.e. $1/6$ and
$1/9$). Thus, providing thermal fluctations do not destroy the
long-ran\-ge order, we expect that these phases could be observed
experimentally below $77K$. We are aware of the approximations we have
made so far and, therefore, no more direct comparisons are going to be
made. A discussion of the validity of this  approach to the other linear
chain structures will be presented in the following section.

\section{Discussion and conclusions}
\label{s5}

The main results obtained in this paper fall into two categories, i.e.
\begin{enumerate}
\item Ground states analysis of the effective energy, eqs (\ref{row5}), 
(\ref{row5a}), and (\ref{row5b}) concerning the role of the
long-ran\-ge indirect interaction. The results were obtained numerically
and this was done without any approximations. We have shown that the
interaction modifies in an essential way the ground states as obtained
from the di\-po\-le-di\-po\-le interaction only. The general tendency is
such that by increasing an amplitude of the indirect
interaction, $A$, one reduces the number of  ground states. 
The full description, however, requires  other parameters of 
the interaction, i.e. $k_F$ and $\varphi$. 
A detailed discussion is presented in section~\ref{s3s3}.
\item Temperature versus coverage (or $T$ vs $\mu$) phase diagrams based
on the effective Hamiltonian ${\cal H}_{eff}$, eqs (\ref{row4}), 
(\ref{row4a}), and (\ref{row4b}), and
computed numerically within the mo\-le\-cu\-lar-field approximation.
\end{enumerate}

It seems that (ii) needs more discussion. It is well known that the
mean--field treatment does not account for (thermal) fluctuations
correctly especially in low dimensional systems \cite{ref1,ref26}. As a
consequence, the long-ran\-ge order might be destroyed and a topology 
of phase diagrams and/or the order of various transitions are not
often correct. This has been discussed and demonstrated explicitly
for short-ran\-ge interactions by comparing mean-field results with the
corresponding Monte Carlo simulations, real-spa\-ce or fi\-ni\-te-si\-ze
renormalization group (cf. \cite{ref38,ref39}).

In our opinion, it would be very interesting to study the problem of
fluctuations in the case of long-ran\-ge (also oscillatory) interactions
in "2.5Dimensional" systems \cite{ref25,ref40}. In this context,
our mean-field results can be treated as  zeroth-or\-der approximation.

The role of the indirect interaction has a great impact on the phase
diagrams despite the fact that it is quite weak ($\frac{A}{a_2}
\sim 0.05 eV $).  Numerical computations show in an explicit way that
this interaction is responsible for causing longer periodic phases to 
develop  high thermal stability. Figs 3-5 illustrate this situation.
The above conjecture, although based on the mean-field theory, is
corroborated by the recent Monte Carlo simulation \cite{ref19}. Note,
however, that the indirect interaction used in ref.~\cite{ref19} is
characterized by only three parameters and has
{\em no-os\-cil\-la\-to\-ry} behaviour.

We think, it is a challange to extend (if practically possible) Monte
Carlo simulation or other more "exact" methods to long-ran\-ge
(oscillatory)
interactions, which seems to be important in adsorption at crystal
surfaces. Also, it would be very interesting to determine experimentally
the whole ($T, \theta$) phase diagrams to check theoretical predictions.

The present work was concerned with adsorption of lithium atoms
on furrowed surfaces of  Mo  and  W . However, other alkaline 
(al\-ka\-li\-ne-earth or rare-earth) atoms are larger in diameter and, 
therefore, our theory will have to be modified for "hard-core" effects.
A kind of modification is also needed to account for more complex 
structures.

This paper has dealt with the indirect interaction decaying like
$\frac{1}{r}$ which is closely  related to both the face and form of the
Fermi surface of the substrate (W and Mo). In general, however, the
indirect interaction has an asymptotic form
$\sim cos(2k_Fr+\varphi)/r^{\beta}$, with $\beta \in [1,5]$ \cite{ref4}.

\noindent We would like to end with the following comment. It is now
well known that the precise power law behaviour of this interaction
might depend upon the form of the Fermi surface, surface electronic
states, virtual elastic distortion of the substrate surface, {\it etc}.
Therefore, it would be very interesting to study a competition between
the dipole--dipole and indirect ($\beta>1$) interactions in view of
possible physical applications. This problem turns out to be quite
difficult for it requires a novel numerical method of  convergence of
series and the results will be published elsewhere.

\subsection*{Acknowledgements}	

The authors are indebted to Professor Jan Ko{\l}aczkiewicz for 
discussions and a critical reading of the manuscript. Several comments
from the referee have been useful. Two of us (J.L and Cz.O) kindly
acknowledge a financial support from the State Committee for Scientific
Research (KBN) via the grant No.~2~P302~113~05. A part of the numerical
computations was performed at the Supercomputing Centre in Pozna\'{n},
Poland.


\end{multicols}

\begin{figure}[h]
\caption{ Coverage as a function of reduced chemical potential,
$\frac{\mu}{\mu_0}$ ($\mu_0=k_BT, T=100 K$), for $\varphi=1.2\pi, 
k_F=0.41 \AA^{-1}$ and $a=3.16 \AA$.
(a) The devil's staircase: $d=1.5$ and $A=0$.
(b) The devil's-like staircase: $d=1.5$ and $A=1$.
(c) The "normal"  staircase: $d=0.5$ and $A=1$.
Here $d$ and $A$ are expressed in units of Debye and $-0.137 eV\AA$,
respectively. For a detailed discussion see the text.
}
\label{dd}
\end{figure}

\begin{figure}[h]
\caption{Mean-field phase diagram: T versus   $\frac{\mu}{\mu_0}$ 
($\mu_0=k_BT, T=100 K$). Here $d=1.5 D$, $E_b=-0.05 eV$, $a=3.14 \AA$.
b and c denote the corresponding insets. For a notation see the text.
}
\label{mf_dd}
\end{figure}

\begin{figure}[h]
\caption{ Mean-field phase diagram (T versus   $\theta$ )
for  $d=1 D$, $A=-0.274 eV\AA$, $k_F=0.47 \AA^{-1}$, $\varphi=0.93\pi$,
$E_b=-0.05 eV$ and $a=3.14 \AA$. For an explanation cf. text.
}
\label{Li_Mo}
\end{figure}

\begin{figure}[h]
\caption{ Mean-field phase diagram (T versus   $\theta$) 
for  $d=1.5 D$, $A=-0.137 eV\AA$, $k_F=0.41 \AA^{-1}$, $\varphi=1.26\pi$,
$E_b=-0.05 eV$ and $a=3.16 \AA$. For an explanation cf. text.
}
\label{Li_W}
\end{figure}

\pagebreak
\begin{table}[tp]
\caption{Sequences of the ground states as a function of phase,
$\varphi$. Here, $k_F=0.41\AA^{-1}$ ($a=3.16 \AA$) and
$k_F=0.47\AA^{-1}$ ($a=3.14 \AA$) are considered because of a possible
application to Li/W(112) and Li/Mo(112). ($\;\frac{d^2}{A}\;$ is
expressed in units of $-7.299 D^2 eV^{-1}\AA^{-1}$).}
\label{tab1}
\begin{tabular}{|c|c|c|c|c|c||c|c|c|c|c|}
\hline 
&
\multicolumn{10}{c|}{$d^{2}/A = 2.25 $} \\
\cline{1-11}
&
\multicolumn{5}{c||}{$k_{F} = 0.41 {\AA}^{-1}$} &
\multicolumn{5}{c|}{$k_{F} = 0.47 {\AA}^{-1}$} \\
\hline
$\varphi/\pi$  & 0.0 & 0.4 & 0.8 & 1.2 & 1.6 & 0.0 & 0.4 & 0.8 & 1.2 & 1.6 \\
\hline \hline
&&&&&&&&&&\\
& 5/14 & 1/11 & 2/23 & 1/15 & 5/14 &  4/22 & 4/22  & 1/14 & 1/13 & 1/5  \\
&& 2/11 & 1/10 & 1/12 &       &  5/22 & 5/22  & 2/23 & 3/22 & 5/20 \\
&& 3/11 & 3/18 & 2/21 &       &  6/22 & 6/22  & 3/22 & 1/6  & 4/15 \\
&& 4/11 & 4/23 & 1/9  &       &  7/22 & 7/22  & 2/11 & 3/15 & 1/3  \\
&& 1/2  & 1/5  & 2/15 &       &  8/22 & 8/22  & 4/19 & 3/11 & 2/5  \\
&       &       & 1/4  & 2/13 &       &  9/22 & 9/22  & 1/4  & 1/3  &\\
&       &       & 8/23 & 4/21 &       & 10/22 & 1/2   & 9/22 & 9/22 &\\
&       &       & 9/23 & 2/10 &       & 11/22 &        & 1/2  & 1/2 & \\
&       &       & 3/7  & 1/4  &&&&&&\\
&       &       & 1/2  & 2/7  &&&&&&\\
&       &       &       & 1/3  &&&&&&\\
&       &       &       & 1/2  &&&&&&\\
&&&&&&&&&&\\
\hline
&
\multicolumn{10}{c|}{$d^{2}/A = 0.25 $} \\
\hline
&&&&&&&&&&\\
& 5/14 & 4/11 & 8/22 & 7/20 & 5/14 & 8/22  & 8/22  & 9/22 & 9/22 & 1/5  \\
&& 1/2  & 1/2  & 2/4  &       & 10/22 & 10/22 & 1/2  &       & 2/5  \\
&       &       &       &       &       & 11/22 & 11/22 &&&\\
&&&&&&&&&&\\
\hline
\end{tabular}

\end{table}

\begin{table}[tp]
\caption{Sequences of the ground states as a function of wavevector, $k_F$,
for  $\varphi=0$ and $a=3.16 \AA$. For a discussion see the text.
($\;\frac{d^2}{A}\;$ is expressed in units of $-7.299 D^2 eV^{-1}\AA^{-1}$).
}
\label{tab2}
\begin{tabular}{|c|c|c|c|c|c|c|c|c|c|c|}
\hline 
&
\multicolumn{10}{c|}{$d^{2}/A = 2.25$} \\
\hline
$k_{F}$  & 0.40 & 0.41 & 0.42 & 0.43 & 0.44 & 0.45 & 0.46 & 0.47 & 0.48 & 0.49 \\
\hline \hline
&&&&&&&&&&\\
& 7/23   & 5/14 & 3/11 & 2/8  & 3/13 &  5/23 & 1/5 & 4/22 & 2/12 &  1/7  \\
& 8/23   && 7/22 & 3/8  & 4/13 &  6/23 & 2/5 & 5/22 & 3/12 &  2/7  \\
& 9/23   && 4/11 & 4/8  & 5/13 &  7/23 &      & 9/22 & 4/12 &  8/21 \\
& 10/23  && 9/22 &       & 6/13 &  8/23 &      &       & 5/12 &  3/7  \\
& 11/23  && 5/11 &       &&  9/23 &      &       & 6/12 & 10/21 \\
&        &&  1/2 &       &       & 10/23 &&&&\\
&        &&      &       &        & 11/23 &&&&\\
\hline
&
\multicolumn{10}{c|}{$d^{2}/A = 0.25 $} \\
\hline
&&&&&&&&&&\\
& 8/23   & 5/14 & 3/11 & 3/8  & 4/13 & 8/23  & 2/5 & 9/22 & 4/12 & 2/7   \\
& 9/23   &       & 7/22 & 4/8  & 6/13 & 9/23  &      &       & 6/12 & 8/21  \\
& 10/23  &       & 9/22 &       &       & 10/23 &&&& 10/21 \\
& 11/23  &       & 5/11 &       &       & 11/23 &&&&\\
&&&&&&&&&&\\
\hline
\end{tabular}
\end{table}
\end{document}